\documentclass[12pt]{iopart}

\usepackage{graphicx}
\usepackage{dcolumn}
\usepackage{bm}

\usepackage{cite}%

\newcommand{\ud}{\mathrm{d}}
\newcommand{\ic}{\mathrm{i}}
\newcommand{\re}{\mathrm{ Re}}
\newcommand{\im}{\mathrm{ Im}}

\expandafter\let\csname equation*\endcsname\relax 

\expandafter\let\csname endequation*\endcsname\relax 

\usepackage{amsmath}
\usepackage{amssymb}

\usepackage{ulem}
\usepackage{color}

\newcommand{\new}[1]{\textcolor{black}{#1}}
\newcommand{\old}[1]{}

\begin{document}

\title{Scattering theory of walking droplets in the presence of obstacles}

\author{R\'emy Dubertrand, Maxime Hubert, Peter Schlagheck, Nicolas Vandewalle, Thierry Bastin, John Martin$^{1}$}
\address{
$^1$  D\'epartement de Physique, CESAM, University of Li\`ege, 4000 Li\`ege,
Belgium}
\ead{\mailto{remy.dubertrand@ulg.ac.be}}

\begin{abstract}
We aim to describe a droplet bouncing on a vibrating bath \new{using a simple and highly versatile model
inspired from quantum mechanics.} Close to the Faraday instability, a long-lived surface wave is created at each bounce, which serves as a pilot wave for the droplet.
This leads to so called walking droplets or walkers.
Since the seminal experiment by {\it Couder et al} [Phys. Rev. Lett. {\bf 97}, 154101 (2006)] there have been many attempts to accurately reproduce the experimental results. \old{Here we present a simple and highly versatile model inspired from quantum mechanics.} We propose to describe the trajectories of a walker using a Green function approach. The Green function is related to the Helmholtz equation with Neumann boundary conditions on the obstacle(s) and outgoing boundary conditions at infinity.
For a single\new{-}slit geometry our model is exactly solvable and reproduces some general features observed experimentally. It stands for a promising candidate to account for the presence of arbitrary boundaries in the walker's dynamics.
\end{abstract}

\maketitle

\section{Introduction}

A considerable attention has recently been paid to the study of a hydrodynamic analogue of quantum wave-particle duality. It started originally from 
an experiment where an oil droplet is falling on a vertically vibrating bath \cite{Nature2005,Couder2005,Couder2006}. In the appropriate regime of 
viscosity and vibrating frequency the drop starts bouncing periodically on the surface. This leads to nontrivial effects due to the coupling between 
the dynamics of the surface wave and the drop, see e.g.~\cite{Protiere2006}. When appropriately tuning the vibrating frequency, the droplet starts 
to move horizontally. This is referred to as a "walking droplet" or "walker". While this walk is rectilinear and {takes place} at constant speed in 
a homogeneous tank, it becomes significantly perturbed in the vicinity of boundaries. In the pioneering experiment \cite{Nature2005} individual 
droplets were walking through a single or double slit. Measuring the droplet positions at a large distance behind the slit(s) yielded similar 
single-slit diffraction and double\new{-}slit interference patterns as in quantum mechanics. In subsequent experiments other quantum phenomena could be 
mimicked like tunnelling \cite{eddi2009}, orbit quantisation and Landau levels \cite{FortPNAS}.

To our knowledge such walking droplets stand for the very first example of systems outside the quantum world that can reproduce some features of pilot wave theory \cite{deBroglie}. Indeed, the droplet can be identified with a particle that creates a wave at each bounce. The surface wave has a back action on the droplet when the latter impacts it, hence acting like a \textit{pilot wave}. This phenomenon is strongly reminiscent of de Broglie's early formulation of quantum theory \cite{deBroglie}, later pursued by Bohm \cite{Bohm}.

A more quantitative comparison between walking droplets and quantum particles has been the motivation of many recent studies, see e.g.~\cite{molacekbush,andersen2015,Brazil2015,Bushreview2015,Batelaan2016} involving various degrees of sophistication in the theoretical modelling. 
The straight-line motion of a droplet on a free surface can be well described by an empirical ansatz which was proposed in \cite{FortPNAS}\new{, see also \cite{LaboussePhD},} and essentially confirmed later in \cite{molacekbush} from a more fundamental perspective\new{, see also \cite{Oza2013} for a stability analysis}. In this approach, the surface wave of the liquid is constituted by a superposition of slowly decaying radial wave profiles that are centred around the previous impact points of the droplet. The local slope of the surface wave then determines the horizontal acceleration of the droplet which\old{ with}\new{, }in combination with various damping and friction effects\new{,} gives rise to an equilibrium speed of the walker. 

While this theoretical approach appears to give a satisfactory description of the behaviour of free walkers, it cannot be applied in the presence of boundaries or obstacles within the liquid, which render the surface wave profile induced by a bounce of a droplet more complicated. The main aim of this paper is to elaborate a straightforward generalisation of the above model in order to incorporate, in principle, arbitrary boundaries and obstacles within the liquid. 
\new{In particular the suggested model keeps the same relevance as the models introduced in \cite{FortPNAS} and \cite{molacekbush}.}
We propose to replace the radially symmetric surface wave profile by the Green function of the Helmholtz operator that properly accounts for the boundary and obstacle under consideration. We assume for this purpose that the surface wave of the liquid exhibit homogeneous boundary conditions (e.g. of Dirichlet, Neumann or Robin type) that render the Helmholtz operator Hermitian. An analytical solution of this model can be obtained for the specific case of a single-slit scattering geometry and in the presence of Neumann boundary conditions, which allows us to express the Green function in terms of series of Mathieu functions.

The paper is organised as follows.
In Sect.~\ref{model} we present our Green function model for the dynamics of a walker in the presence of a boundary. 
In Sect.~\ref{single_slit} we apply our model to the specific case of a single-slit geometry for which an analytical expression can be obtained for the Green function. The resulting surface wave profile is then used to numerically propagate droplets across the slit for various impact parameters and thereby obtain a theoretical prediction for the diffraction pattern. In Sect.~\ref{conclusion} we discuss the benefits and the limitations of our model and expose possible extensions of it.

\section{Green function approach for walking droplets}
\label{model} 

\subsection{Walkers in free space}


It is useful to recall the models that were previously used in the absence of obstacles \cite{Protiere2006}. The starting assumption is that, when a droplet impacts the surface wave, it creates a perturbation of small amplitude so that the equations describing the bath surface can be linearised. 
Then it is customary to decompose the motion of the walker in the directions along and transverse to the vertical vibrating direction. The first refers to the bouncing and can be approximated to be periodic, if the wave amplitude at the surface is small enough \cite{Protiere2006}. A recent study \old{claimed}\new{demonstrated} that the vertical bouncing can become chaotic depending on the size of the droplet and the distance to the Faraday instability threshold \cite{chaotic_bouncing}. For sake of simplicity we will assume that it remains periodic.

The horizontal motion of the droplet will be our main focus.
Let us denote by the $2-$dimensional vector ${\bf r}(t)$ the position of the droplet's impact on the interface between liquid and air at time $t$. We want to write a dynamical equation for ${\bf r}(t)$.
The historically first model \cite{Protiere2006} assumes that the droplet is a material point as in classical mechanics. 
It is subject to three types of forces:
\begin{itemize}
\item a force originating from the coupling between the surface wave of the bath and the droplet. This coupling is taken to be of the form $-A {\bf \nabla} h({\bf r},t)$, where $h({\bf r},t)$ is the height of the fluid surface at the position ${\bf r}$ and time $t$ and $A$ is a coupling coefficient to be discussed below,
\item a friction force due to the viscosity of the air layer when the droplet {skids on} during the contact time. At the leading order of small velocities it is modelled by $-D\ud {\bf r}/\ud t$ where the coefficient $D$ depends on the mass and the size of the drop\old{;}\new{,} as well as on the density, the viscosity and the surface tension of the fluid \cite{molacekbush}. It should be noted that other forms of dissipation 
have been discussed in \cite{molacekbush}. Note that they always have the form mentioned above, i.e.~proportional to the velocity vector,
\item any external force ${\bf F}_{\rm ext}$ applied to the droplet. For example droplets with metallic core have been designed and put in a magnetic field in order to create a harmonic potential \cite{Perrard2014}.
\end{itemize}
Under these assumptions we can write a Newton-like law for the droplet's horizontal dynamics
\begin{equation}
  \label{Newton_drop}
  m \frac{\ud^2 {\bf r}}{\ud t^2}={\bf F}_{\rm ext}-D\frac{\ud {\bf r}}{\ud t} - A {\bf \nabla} h({\bf r},t)\ ,
\end{equation}
where $m$ denotes the mass of the droplet. In the present study, and for the sake of simplicity, we will assume that ${\bf F}_{\rm ext}={ 0}$. 

The last and highly nontrivial part of the model is the ansatz for the surface of the fluid. {In} our opinion this is the very source of all the complexity of the droplet's dynamics. {The} first ansatz for the fluid surface has been proposed in \cite{FortPNAS}. It reads:
\begin{equation}
  \label{h_Fort}
  h({\bf r},t_n)=\sum_{p=-\infty}^{n-1} \re\left[\dfrac{C_0 e^{ \ic  k_F|{\bf r}-{\bf r}_p|+\ic \phi} }{|{\bf r}-{\bf r}_p|^{1/2}} \right] e^{-|{\bf r}-{\bf r}_p|/\delta} e^{-\frac{n-p}{{\cal M}} } \ ,
\end{equation}
where \old{${\bf k}_F$}\new{$k_F$} is the wave \old{vector}{\new{number} of the Faraday waves created by the bath vibration. The parameters $C_0$ and $\phi$ are the intrinsic amplitude and phase of emitted Faraday waves, which can be estimated experimentally, see e.g.\;\cite{eddi2011}. The spatial damping term comes from the viscosity: $\delta$ is the typical length scale that a wave can travel at the fluid surface. It has been estimated from experimental data in \cite{eddi2011} although the details of the determination of $\delta$ were not {made explicit}. The vector ${\bf r}_p$ stands for the impact position of the droplet at time $t_p\equiv p T_F < t_n$, where $T_F$ is the period of Faraday waves. Notice that the index $n$ for the time recalls that we are interested in the surface profile only at a discrete sequence of times, when the droplet interacts with it.
Eventually Faraday waves are subject to a temporal damping, which is characterised by the key parameter $\cal M$, often called the memory. $\cal M$ is related to the difference between the vibration amplitude $\Gamma$ of the bath and the Faraday threshold $\Gamma_F$
\begin{equation}
  \label{def_mem}
  {\cal M}=\dfrac{\Gamma_F}{\Gamma_F-\Gamma} \ .
\end{equation}
Physically, as the vibration amplitude is always below the threshold in the walking regime, this means that a perturbation of the surface profile will lead to a Faraday wave, which will last typically for the duration ${\cal M} T_F$. 

Another {expression} was derived for the surface height from a more fundamental perspective \cite{molacekbush}. When there is no obstacle, the surface height can be modelled by:
\begin{equation}
  \label{h_Molacek}
  h({\bf r},t_n)=h_0 \sum_{p=-\infty}^{n-1} J_0(k_F|{\bf r}-{\bf r}_p|) e^{-\frac{n-p}{{\cal M}} }\ ,
\end{equation}
where $h_0$ is a function of the fluid and droplet parameters. $J_0$ denotes the Bessel function of the first kind of zeroth order. 

While there are obvious similarities between both ansatz (\ref{h_Fort}) and (\ref{h_Molacek}), we want to comment important differences. The main viscosity effects in (\ref{h_Molacek}) are located in $h_0$ (there is no spatial damping). 
Equation (\ref{h_Molacek}) offers a smoother spatial profile at the vicinity of the impact, while it behaves in the same way (the amplitude decreasing like $1/\sqrt{r}$) as Eq.~(\ref{h_Fort}) at larger distances.
This model can be generalised by taking a time continuum limit \cite{Brazil2015}.
\subsection{Obstacles for walkers} 
\label{obsta}

It is worth recalling that in the previous experiments \cite{Couder2006,Filoux2015a,Filoux2015b} an obstacle consists {of} a submerged {barrier}. This changes the local depth of the bath and hence the dispersion relation for the Faraday waves. 
{For a sufficiently small depth these waves are so strongly damped that this effectively leads to a region into which the walking droplets generally cannot go (except for occasional "tunneling" events \cite{eddi2009}) even though surface waves may slightly penetrate this region (as was observed using the free surface schlieren technique \cite{eddi2009}).} With this definition of a boundary there have been several geometries considered to study the dynamics of a walker: the circular cavity \cite{harris2013}, the annular cavity \cite{Filoux2015b}, the square cavity \cite{eddi2009,Shirokoff2013,Gilet2014} and a droplet in a rotating tank \cite{FortPNAS,oza2014}.  One should emphasise that one of the most intriguing results obtained with the walkers has been encountered within the single\new{-} and double\new{-}slit geometries, where an interference pattern was experimentally observed \cite{Couder2006}.

On the theoretical side, the presence of obstacles seems to resist a systematic treatment. Secondary sources were suggested in \cite{Couder2006} with poor physical justification. 
A recent study {has} focused on the circular cavity \cite{Tristan2015}. It relies on a decomposition of the surface wave into the eigenmodes of the cavity. In this model the surface wave is assumed to obey {zero} Neumann conditions at the boundary, i.e., the normal derivative of the modes vanishes at the boundary. So far this model only deals with confined geometries.

\subsection{Our model: Green function approach}

We choose to adopt here a conceptually simpler and more direct approach. The main goal is to account for any geometry of the tank as well as for any shape of one or several obstacles inside it. As usual in fluid dynamics, the main problem is to describe precisely the boundary conditions. To this end, we recall that in the vanishing viscosity limit the Faraday waves can be described by imposing {zero} Neumann boundary conditions \cite{BenjaminUrsell}.
As the small viscosity approach was already successfully applied to describe the walking droplet, we choose to assume that the {surface} waves should obey these boundary conditions along the boundary of any obstacle. Generalisations to other {homogeneous boundary conditions (Dirichlet or Robin)} {are} straightforward.

Our model then relies on the Newton-like description of the droplet via Eq.~(\ref{Newton_drop}) for the horizontal motion as it has been the approach which {best agrees} with the experimental data. 
The starting point is to notice that the {expression} (\ref{h_Molacek}) for the bath surface without any boundary can be rewritten as
\begin{equation}
  \label{bathsurf_ansatz}
  h({\bf r},t_n)=-4 h_0 \sum_{p=-\infty}^{n-1} \im \left[G_0({\bf r},{\bf r}_p)\right]  e^{-\frac{n-p}{{\cal M}} }\ ,
\end{equation}
where the Green function for the Helmholtz equation in the $2-$dimensional Euclidean plane has been introduced:
\begin{equation}
  \label{Greenfree2D}
  G_0({\bf r},{\bf r}_0)=\frac{H_0^{(1)}(k_F |{\bf r}-{\bf r}_0|)}{4\ic}\ .
\end{equation}
Here $H_0^{(1)}(z)=J_0(z)+\ic Y_0(z)$ denotes the Hankel function of the first kind of order $0$. 

Our model consists {of} generalising Eq.~(\ref{bathsurf_ansatz}) in the presence of obstacles by considering the relevant Green function. More precisely, the bath surface will be described by:
\begin{equation}
  \label{ansatz}
  h({\bf r},t_n)=-4 h_0 \sum_{p=-\infty}^{n-1} \im \left[G({\bf r},{\bf r}_p)\right]  e^{-\frac{n-p}{{\cal M}} }\ ,
\end{equation}
where $G({\bf r},{\bf r}_0)$ is the kernel of a certain Green operator. It is defined through the following requirements:
\begin{itemize}
\item $G({\bf r},{\bf r}_0)$ is the Green function for the Helmholtz equation with the wave number $k_F$:
\begin{equation}
  \label{Green_eq}
  ({\bf \nabla}^2+k_F^2)G({\bf r},{\bf r}_0)=\delta({\bf r}-{\bf r}_0)\;,
\end{equation}
where $\delta({\bf r})$ stands for the Dirac distribution, and ${\bf r}_0$ usually refers to a source (see below),
\item it obeys Neumann boundary conditions on the obstacles, see Sect.~\ref{obsta},
\item it obeys outgoing boundary conditions at infinity:
\begin{equation}
  G({\bf r},{\bf r}_0)\propto \frac{e^{+\ic k_F r}}{\sqrt{ k_F r}}
,\quad r\to \infty\ .\label{outgoingBC}
\end{equation}
\end{itemize}
The model containing Eq.~(\ref{ansatz}) for the bath surface together with the above listed requirements in order to uniquely define $G({\bf r},{\bf r}_0)$ constitute the main ingredients of the present theory. 
 {For sake of completeness it is assumed that a small positive imaginary part has to be added to $k_F^2$ in (\ref{Green_eq}) so that $G({\bf r},{\bf r}_0)$ stands for the \textit{retarded} Green function.}

We will now explain why the imaginary part of $G({\bf r},{\bf r}_0)$ is relevant for our model.
When a droplet of infinitesimal spatial extent hits the bath at the point ${\bf r}_0$, one can model the bath surface receiving one point impact by
\begin{equation}
\label{expand_h0}
  h_p({\bf r}) \propto \delta({\bf r}-{\bf r}_0)
\end{equation}
{
We make the general assumption that there exists a continuous eigenbasis of the Laplacian operator $\nabla^2$ consisting of smooth, real, orthogonal and properly normalised functions $\varphi_{k,\ell}({\bf r})$ that satisfy the boundary conditions on the obstacles. These functions are parameterized by an continuous index $k$ that represents the wave number related to the eigenvalues $-k^2$ of the Laplacian as well as a discrete index $\ell$ that accounts for the degeneracy of the eigenspectrum. 
This set of functions $\varphi_{k,\ell}({\bf r})$ is assumed to be complete (i.e., there is no bound state) and one can decompose (\ref{expand_h0}):
\begin{equation}
  h_p({\bf r}) \propto \delta({\bf r}-{\bf r}_0)=\sum_{\ell}\int_0^\infty \varphi_{k,\ell}({\bf r}_0) \varphi_{k,\ell}({\bf r})\ud {k}\ .
\label{expand_hp}
\end{equation}
}

We assume that after one bounce the capillary waves emitted by the impact of the droplet has entirely left the impact region of the droplet\footnote{This argument needs to be refined when the impact occurs very close to the boundary. 
Such impacts, however, are not expected to occur often along the trajectory of a droplet.}. The surface profile is then dominantly governed by standing Faraday waves. This is in agreement with the observations reported in \cite{eddi2011}.
Consequently, we now assume that only those components of the decomposition of \old{$h_0$ in Eq.~(\ref{expand_h0})}
\new{$h_p$ in Eq.~(\ref{expand_hp})} survive, whose wave number (or, more precisely, the eigenvalue of Helmholtz equation) is identical to the Faraday wave number $k_F$. A more detailed description of the decomposition between capillary and Faraday waves will be provided in a forthcoming publication.
This yields the expression for the surface profile at the next bounce of the droplet:
\begin{equation}
\label{profile_t}
  h_{p+1}({\bf r}) \propto 
\sum_{\ell} \int_0^\infty \varphi_{ k,\ell}({\bf r}_0) \varphi_{k,\ell}({\bf r})\delta(k-k_F)\ud {k}\ ,
\end{equation}
where the proportionality factor accounts for temporal decay due to the memory.
The decomposition of the retarded Green function defined in (\ref{Green_eq}) into eigenstates is
\begin{equation}
  \label{expand_G_general}
  G({\bf r},{\bf r}_0)=\lim_{\epsilon\to 0^+}\sum_{\ell}\int_0^\infty \dfrac{\varphi_{k,\ell}({\bf r_0}) \varphi_{k,\ell}({\bf r})}{k_F^2-{k}^2+\ic\epsilon}\ud { k}
\end{equation}
As mentioned above a small real number $\epsilon$ was added to fix the prescription of the Green function as we chose the retarded one. Next the following identity for distributions is used:
\begin{equation*}
  \lim_{\epsilon\to 0^+}\dfrac{1}{x+\ic \epsilon}={\rm PV}\frac{1}{x} -\ic \pi \delta(x)\ ,
\end{equation*}
where ${\rm PV}$ refers to the Cauchy principal value. This allows us to obtain the surface profile formed by one bounce:
\begin{equation}
  \label{h_0_G}
  h_{\textrm{1 bounce}}({\bf r}) \propto \im\left[ G({\bf r},{\bf r}_0)\right]\ .
\end{equation}
The final expression in Eq.~(\ref{ansatz}) comes from a superposition argument: the resulting surface profile is the sum of all the Faraday waves emitted during by the previous impacts of the droplet.

It is worth giving some remarks about our model. First, it reproduces the dynamics of a walker as derived from fluid dynamics arguments in \cite{molacekbush} when there is no obstacle. The description using Helmholtz equation for the bath profile is also used in \cite{Tristan2015}, but in a different manner: the surface profile is expanded as a superposition of eigenmodes of the cavity. Our approach is similar to this idea. It is more general as it applies to both closed and open geometries.

Furthermore, our model allows us to draw a straightforward analogy with the quantum mechanics of a two-dimensional particle. In free space, the latter is described by the Schr\"odinger equation:
\begin{equation}
  \label{Schro}
  \ic\hbar\dfrac{\partial}{\partial t} \Psi({\bf r},t)=-\frac{\hbar^2}{2m}{\bf \nabla}^2 \Psi({\bf r},t)\ ,
\end{equation}
where $\hbar=h/2\pi$ is the reduced Planck constant and $m$ is the mass of the particle. The presence of obstacles can be incorporated by adding a potential term to Eq.~(\ref{Schro}) or by defining appropriate boundary conditions (which would, most generally, be of Robin type) at the borders of the obstacles. The scattering process of a wave packet that is launched towards a specific geometry of obstacles can then be represented by a coherent superposition of waves that are described by the retarded Green function $G({\bf r},{\bf r}_0)\equiv G({\bf r},{\bf r}_0;E)$ of the Helmholtz operator satisfying the appropriate boundary conditions at the obstacles with the associated energy $E={\hbar^2 k_F^2}/{2m}$. Up to a constant prefactor, we obtain
\begin{equation}
   \Psi({\bf r},t)\propto \int \ud E \int \ud {\bf r}_0\ G({\bf r},{\bf r}_0;E) e^{-\ic Et/\hbar} \Psi({\bf r}_0,0)\ ,
\end{equation}
where $\Psi({\bf r}_0,0)$ represents the initial state of the wave packet. We note that an analogy with Schr\"odinger equation has also been suggested in \cite{andersen2015} where a Bohmian like model was use to compute the trajectories of a walker.

\section{Walkers going through a single slit}
\label{single_slit}

We will illustrate the model introduced in the previous section by considering a special choice for the obstacle. More precisely, the trajectories of walkers going through a single slit are considered. This obstacle is motivated by two main reasons: first, it was among the first geometries to be considered in the experiments \cite{Couder2006}. Second, it is among the few shapes {for which} an explicit and analytical expression of the Green function {can be derived}.

\subsection{Green function of the single slit}

In order to write the Green function of Eq.~(\ref{Green_eq}) with Neumann boundary conditions on a single slit, it is convenient to introduce the elliptic coordinates $(u,v)$ in a $2D$ geometry:
\begin{eqnarray}
 x&=& \frac{a}{2} \cosh u\cos v\ ,\label{eqcoordell1}\\
y&=&\frac{a}{2} \sinh u\sin v\ , \label{eqcoordell2}
\end{eqnarray}
where $(x,y)$ are the Cartesian coordinates. The range of the new coordinates is:
$$ u\ge 0,\ -\pi< v\le \pi\ .$$
In this definition of the elliptic coordinates, $a$ denotes the width of the slit and the arms of the slit are along the $x$ axis,  see also Fig.~\ref{ellcoord}.
\begin{figure}[!ht]
  \centering
  \includegraphics[width=.6\textwidth]{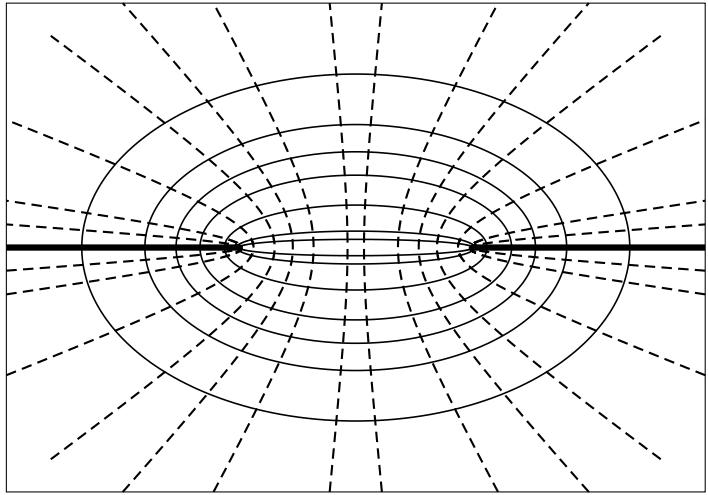}
  \caption{Elliptic coordinates as defined in (\ref{eqcoordell1}), and (\ref{eqcoordell2}). The full lines are the $u=u_0$ level curves. 
The dashed lines are the $v=v_0$ level curves. 
The two segments of thicker lines stand for both arms of the single slit. The Cartesian coordinates of the finite end of each arm are $(-a/2,0)$, and $(a/2,0)$.}
  \label{ellcoord}
\end{figure} 
The elliptic coordinates are very convenient for the single\new{-}slit problem because both arms have very simple equations: the left arm in Fig.~\ref{ellcoord} is defined as $v=\pi$, whereas the right arm is $v=0$. With our {definition} of $v$, the upper half plane is $u>0,v>0$ while the lower half plane is $u>0,v<0$.
The slit is described by $u=0$. Along the slit, the points with coordinates $(u,v)=(0,v)$ and $(u,v)=(0,-v)$ coincide for $0< v < \pi$.

There have been several studies for the derivation of the Green function for the single slit \cite{schwarzschild,sieger,strutt}. 
The technical details of its evaluation are beyond the scope of this paper and are to be published elsewhere.
A reminder of the derivation is given in \ref{Green}. In the following the point ${\bf r}$ in the plane is assumed to have $(u,v)$ as elliptical coordinates while ${\bf r}_0$ is identified with $(u_0,v_0)$.
Without loss of generality one can consider that the source is located below the slit, i.e. $v_0<0$. Then the Green function for the single slit with Neumann boundary conditions is in the upper half plane ($0<v< \pi$):
\begin{equation}
  \label{Gsingleslit_up}
  G({\bf r},{\bf r}_0)=
\displaystyle\sum_{n\ge 0} \dfrac{Me^{(1)}_n(q,u)Me^{(1)}_n(q,u_0)ce_n(q,v)ce_n(q,v_0)}{\pi Me^{(1)}_n(q,0)Me^{(1)\;\prime}_n(q,0)}
\end{equation}
and in the lower half plane ($-\pi < v < 0$):
\begin{equation}
  \label{Gsingleslit_down}
 G({\bf r},{\bf r}_0)=\displaystyle\sum_{n\ge 0} \left[2\dfrac{Me^{(1)}_n(q,u_>)Ce_n(q,u_<)}{\pi ce_n(q,0) } -
 \dfrac{Me^{(1)}_n(q,u)Me^{(1)}_n(q,u_0)}{\pi Me^{(1)}_n(q,0)}\right]\dfrac{ce_n(q,v_0) ce_n(q,v)}{Me^{(1)\;\prime}_n(q,0) }\ .
\end{equation}
$ce_n(q,v)$ refers to the even Mathieu functions while $Ce_n(q,u)$ and $Me^{(1)}_n(q,u)$ are solutions of the associated (also known as radial) Mathieu equation, see \ref{Mathieu}. We also introduced the symbols $u_<\equiv{\rm min}(u,u_0)$ and $u_>\equiv{\rm max}(u,u_0)$. The second parameter entering the Mathieu equation is:
\begin{equation}
  \label{defq}
  q=\left(\dfrac{k_F a}{4}\right)^2\ .
\end{equation}
An illustration of the Green function resulting from the expressions (\ref{Gsingleslit_up}) and (\ref{Gsingleslit_down}) is shown in Fig.~\ref{Gsslit}.
\begin{figure}[!ht]
  \centering
  \includegraphics[width=0.8\textwidth]{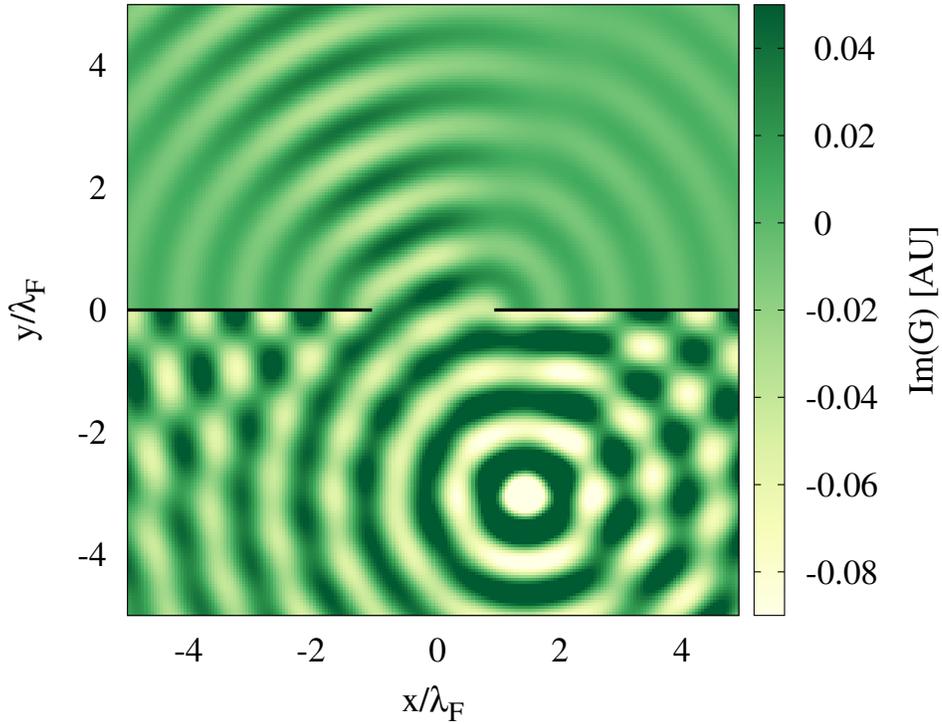}
  \caption{Imaginary part of the Green function as defined in Eq.~(\ref{Gsingleslit_up}) and Eq.~(\ref{Gsingleslit_down}) for $k_F a=4\pi$. The source is located at the point with Cartesian coordinates $(1.5,-3)$. 
  }
  \label{Gsslit}
\end{figure} 

\subsection{Trajectories of walkers in the presence of a single slit}

The fluid parameters are taken such that they {are consistent with} the experimental data for silicon oil with a viscosity $\nu=20$ cSt. The acceleration provided by the shaker is
\begin{equation}
  \label{accel}
  \Gamma(t)=\Gamma\cos\omega_0 t\ ,
\end{equation}
with $\Gamma=4.2g$ and ${\omega_0}/{2\pi}=80$ Hz \cite{Couder2006}.
In our numerical implementation 
the series (\ref{Gsingleslit_up}) and (\ref{Gsingleslit_down}) to compute the Green function have been truncated to $n\le 100$ and the superposition of sources in Eq.~(\ref{ansatz}) has been taken to start at $p=-5{\cal M}+1$. 

We shall study in the following the influence of three parameters on the diffractive character of the trajectories: the memory ${\cal M}$, the inertia via $\tau_v$ (see below) and the slit width $a$. 
Intuitively, the memory parameter is the number of past bounces that the bath remembers. It is a crucial parameter as the diffraction feature of the walker is believed to occur at large memory. Inertia is quantified by the friction when the droplet skids on the liquid. During this surfing phase the droplet is subject to a friction force due to the viscosity of the liquid with a characteristic time $\tau_v$. This time is related to the coefficient $D$ in Eq.~(\ref{Newton_drop}) via 
\begin{equation}
  \label{Dtotau_v}
  \frac{D}{m}=\frac{1}{\tau_v}\ .
\end{equation}
In a more physical perspective this friction is also related to the droplet's mass or its size: the larger the droplet, the larger $\tau_v$. The slit width $a$ will be chosen to be of the order of the surface wave length. Indeed, if a quantum particle is sent to a diffracting aperture like a slit, the diffraction effects will be different if the de Broglie wavelength of the particle is much larger or of the order of the slit width. Therefore varying the slit width $a$ is useful to make \textit{quantitative} comparison between a droplet and a quantum particle.

In Figs.~\ref{sslit_a2_t0.09}, \ref{sslit_a2_t0.0225} and \ref{sslit_a1_t0.0225} the trajectories obtained from the integration of Eq.~(\ref{Newton_drop}) using Eq.~(\ref{ansatz}) are shown for the case of the single slit geometry. These trajectories are obtained from an ensemble of initial positions that are characterised by a fixed distance from the slit and a variable lateral position. The initial velocity of the walker is assumed to be perpendicular to the orientation of the slit such that the walker moves right away to the slit. The magnitude of the initial velocity has been fixed to be $10$ mm.s$^{-1}$. The time between two bounces is taken to be $0.025$ s corresponding to a vertical shaking frequency of $80$ Hz.

First, we investigate the effects of the memory parameter ${\cal M}$ as defined in (\ref{def_mem}) for  $\tau_v=0.09$~s, and $a=2\lambda_F$ in Fig.~\ref{sslit_a2_t0.09}. Results are shown for the values ${\cal M}=10$ and ${\cal M}=30$.
The histogram for the angular distribution in the far field behind the slit is evaluated in the second row of the figure. It was checked that the trajectories are rectilinear for subsequent times.
Each trajectory is stopped when it crosses in the upper half plane the far field circle $|{\bf r}|=15\lambda_F$. 
The final position is stored by computing the angle between the segment joining the final point to the origin and the vertical axis in Cartesian coordinates.
These histograms show an oscillating pattern similar to what was observed in \cite{Couder2006}, but the range of the angular far field directions is much more narrow. 
Furthermore, there is a subtle difference between ${\cal M}=10$  and ${\cal M}=30$. Both values of the memory parameter give rise to
a selection of specific directions in the far field but the oscillation pattern in the histogram has a smaller amplitude in the case of a larger memory. This is consistent with the observation that, in the considered speed regime of the droplet, increasing the memory forces the walker to follow straighter trajectories.

\begin{figure}[!ht]
    \begin{center}
      \includegraphics[width=\linewidth]{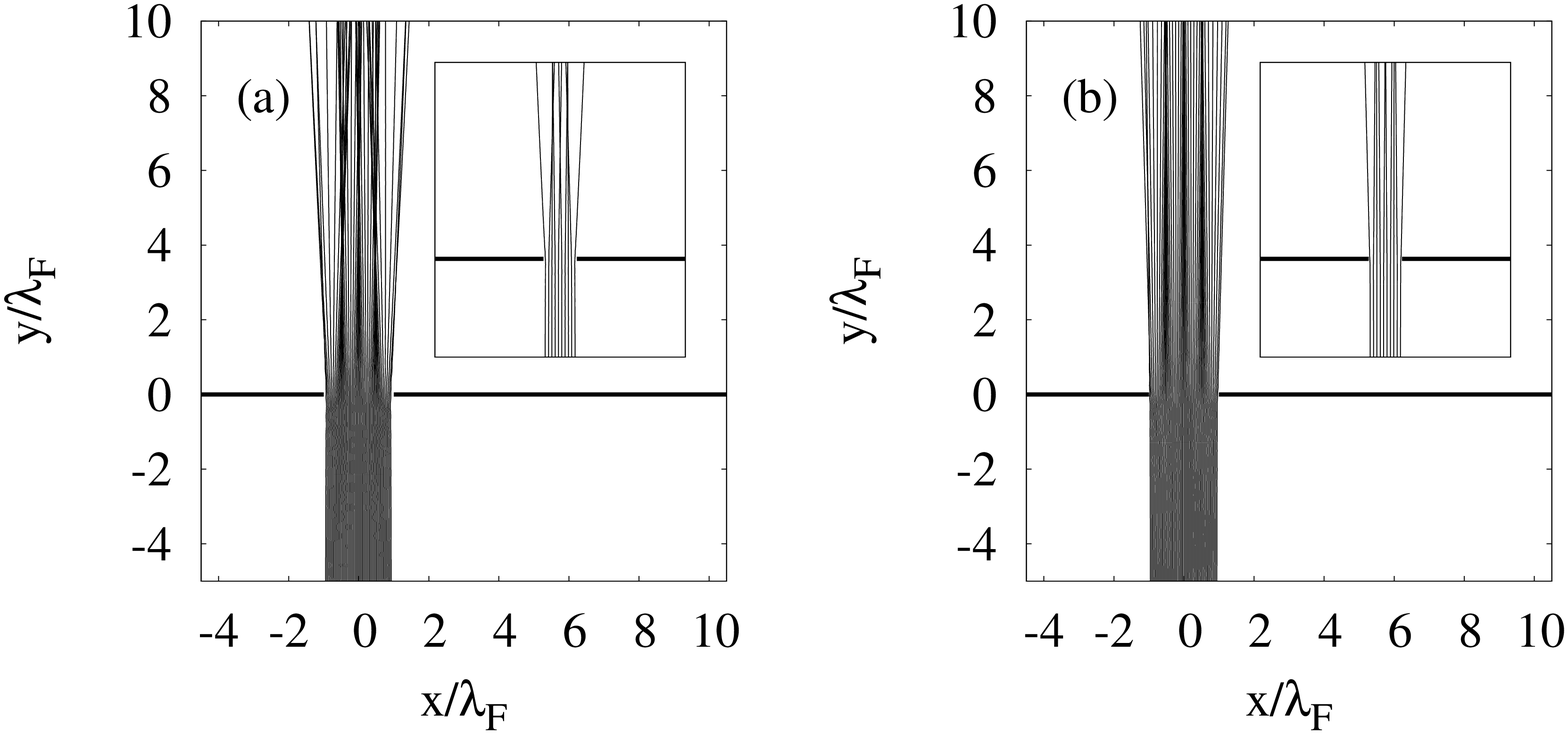}\\
      \includegraphics[width=0.9\linewidth]{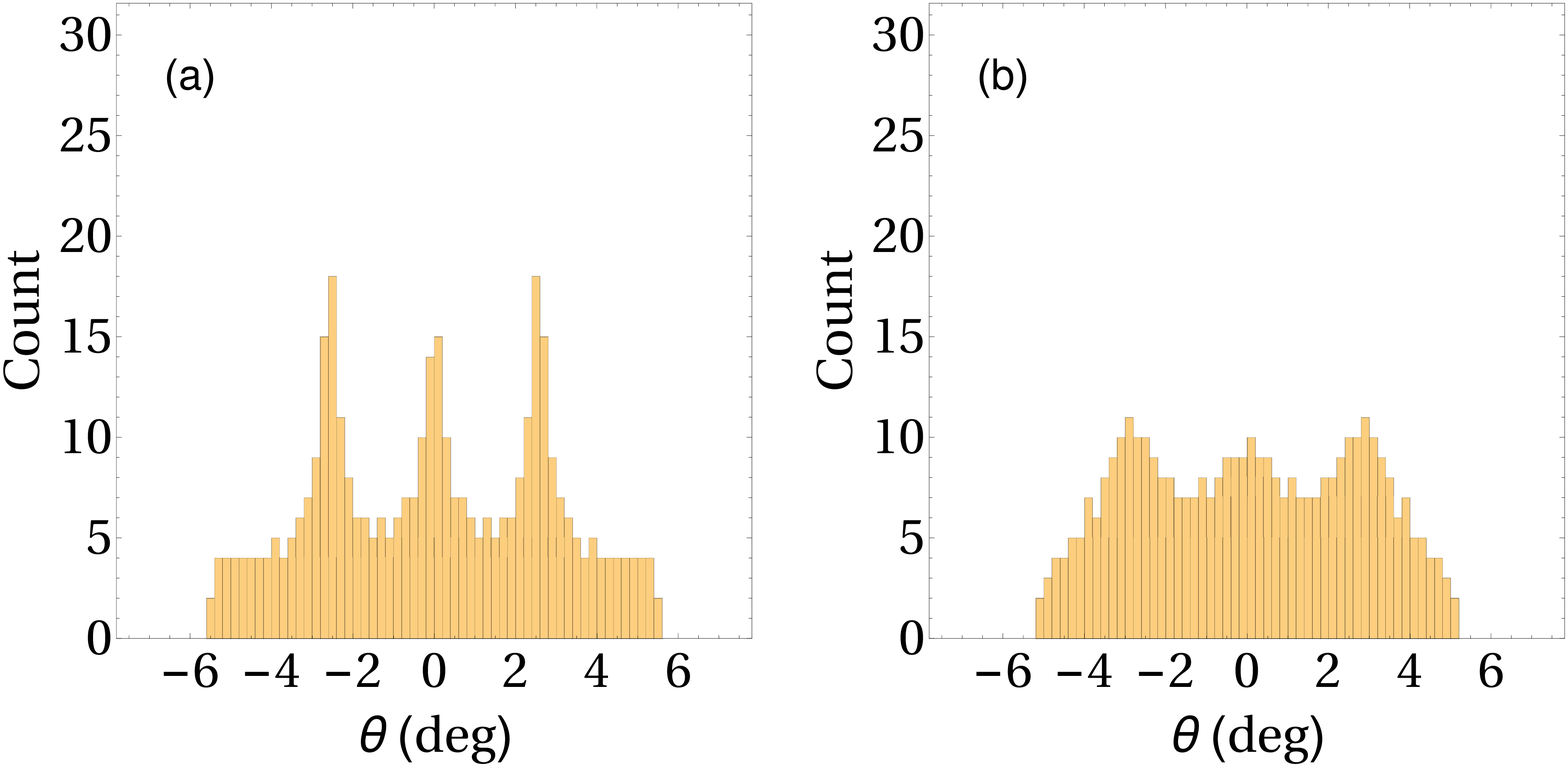}
    \end{center}
  \caption{Top: Droplet trajectories through a single slit computed from the model defined with Eqs.~(\ref{Newton_drop}) and (\ref{ansatz}). $k_F a=4\pi$, and $\tau_v=0.09$~s. Every initial trajectory originates from the segment $|x|\le a=2\lambda_F$ and $y=-5\lambda_F$ with an upward vertical velocity. \new{Each inset shows the same data with less initial points. }Bottom: Histogram of the final positions of the same trajectories as in the top line. The vertical scale gives the absolute number of stored position. a) ${\cal M}=10$. b) ${\cal M}=30$.}
  \label{sslit_a2_t0.09}
\end{figure}

Next, we decide to investigate the {sensitivity} of our results {with respect to} a variation of {the friction time scale} $\tau_v${. This is shown in Fig.~\ref{sslit_a2_t0.0225} where  we consider the same slit width $a=2\lambda_F$ but a shorter friction time scale $\tau_v=0.0225$~s.} We show the results for the same two values of the memory parameter.
\begin{figure}[!ht]
    \begin{center}
      \includegraphics[width=\linewidth]{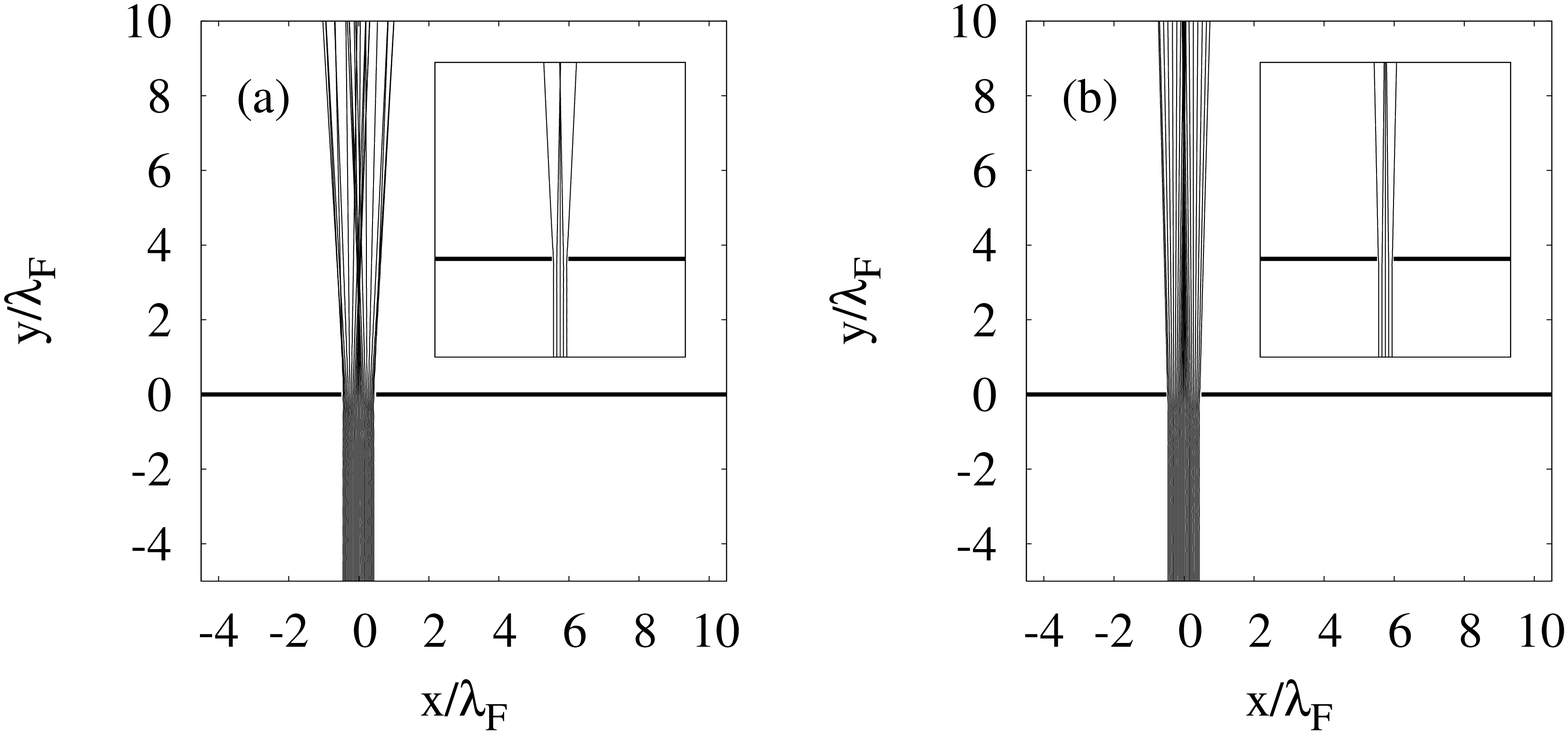}\\
      \includegraphics[width=0.9\linewidth]{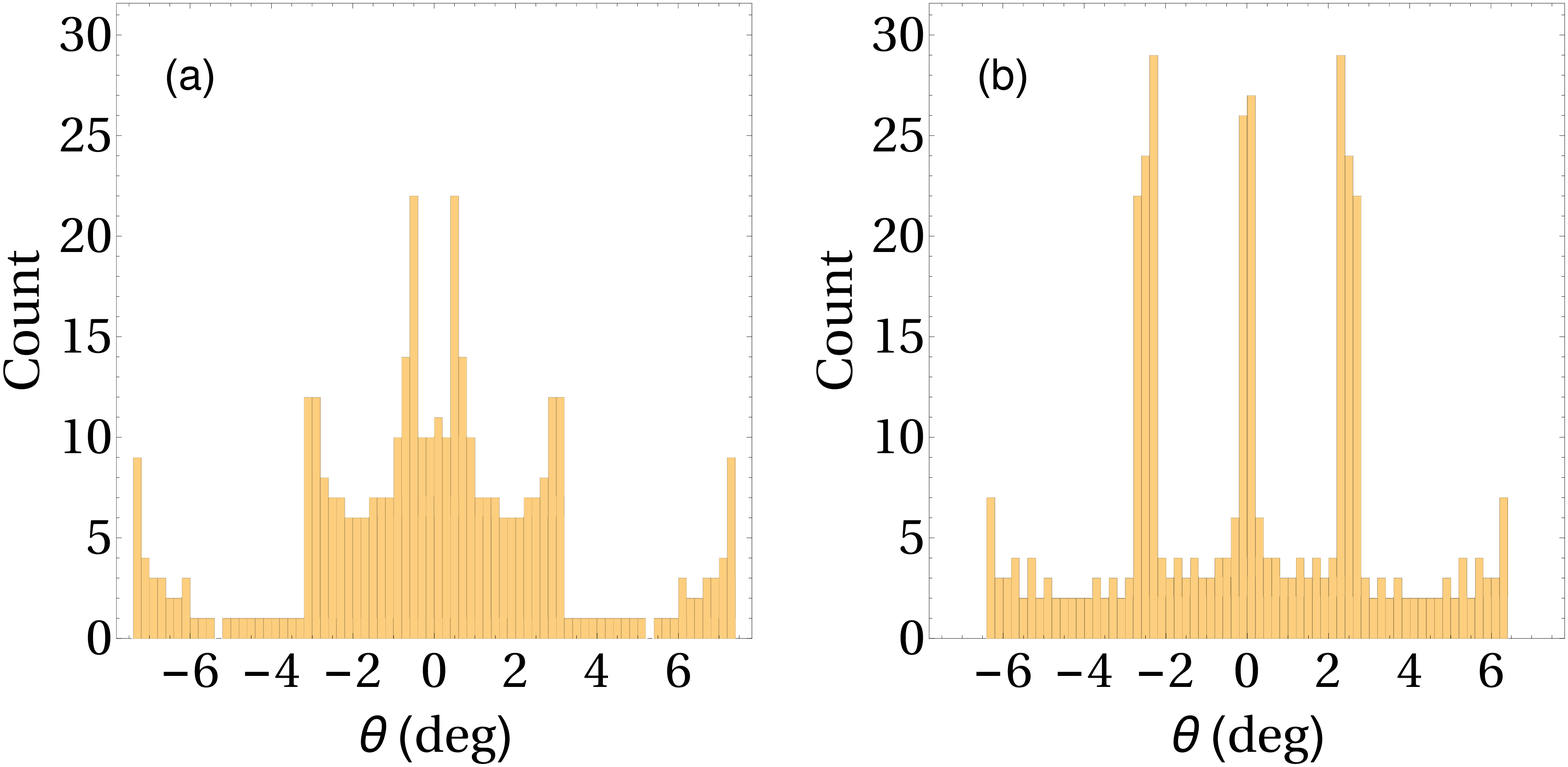}
    \end{center}
  \caption{Same as Fig.~\ref{sslit_a2_t0.09} with $\tau_v=0.0225$~s.}
  \label{sslit_a2_t0.0225}
\end{figure}
{While the range of angular far-field directions is similar to Fig.~\ref{sslit_a2_t0.09}, the far-field pattern of the droplet trajectories significantly depend on the value of $\tau_v$.} {This indicates} that the viscous friction during the surfing phase has a strong influence on the properties of the trajectories. Notice also that the increase of the memory parameter (\ref{def_mem}) leads to a higher angular selection for the trajectories behind the slit. 

Last, we show how the trajectories depend\old{s} on the width of the slit. In Fig.~\ref{sslit_a1_t0.0225} the trajectories and the histogram for the far field direction are shown for $\tau_v=0.0225$~s, and $a=\lambda_F$.
\begin{figure}[!ht]
    \begin{center}
      \includegraphics[width=\linewidth]{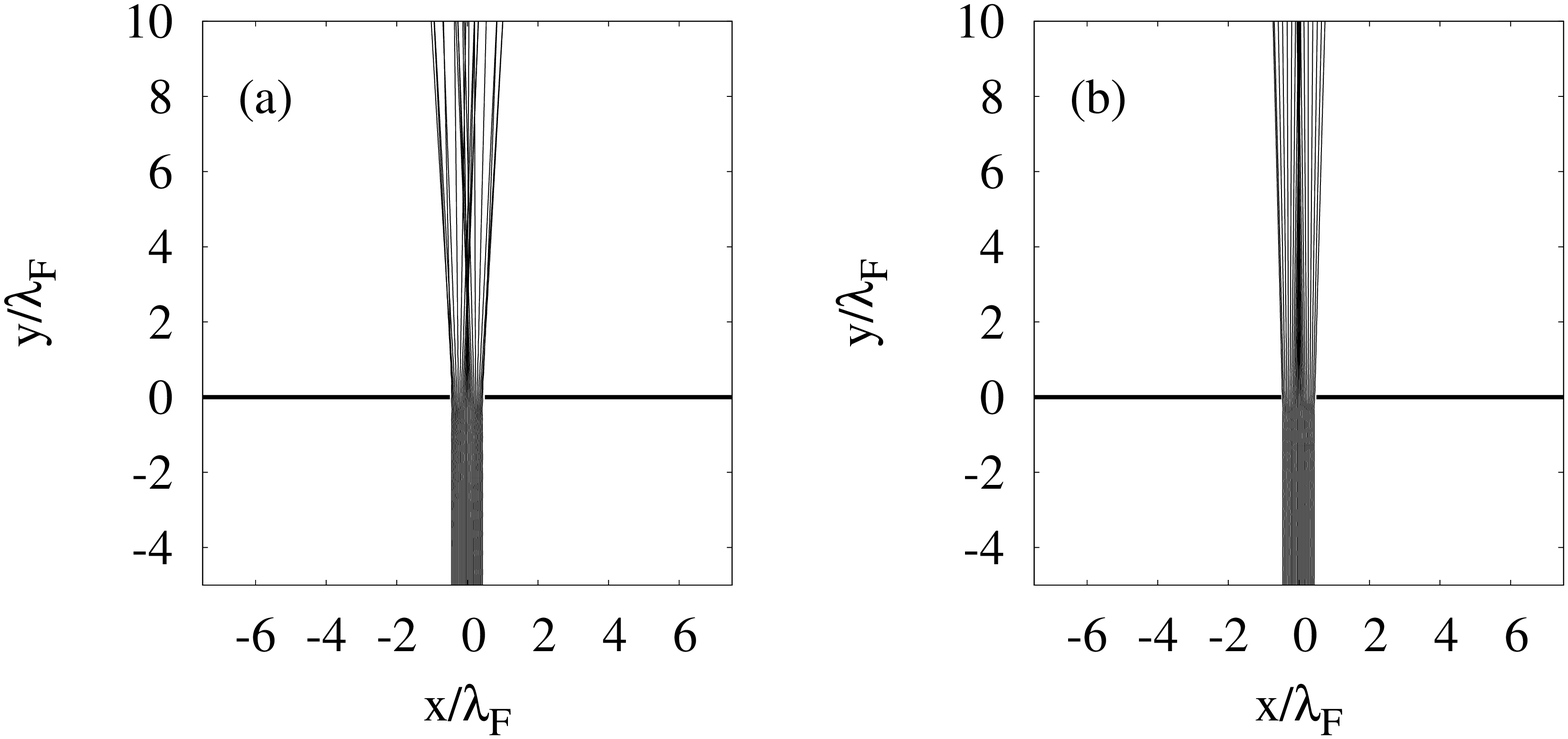}\\
      \includegraphics[width=0.9\linewidth]{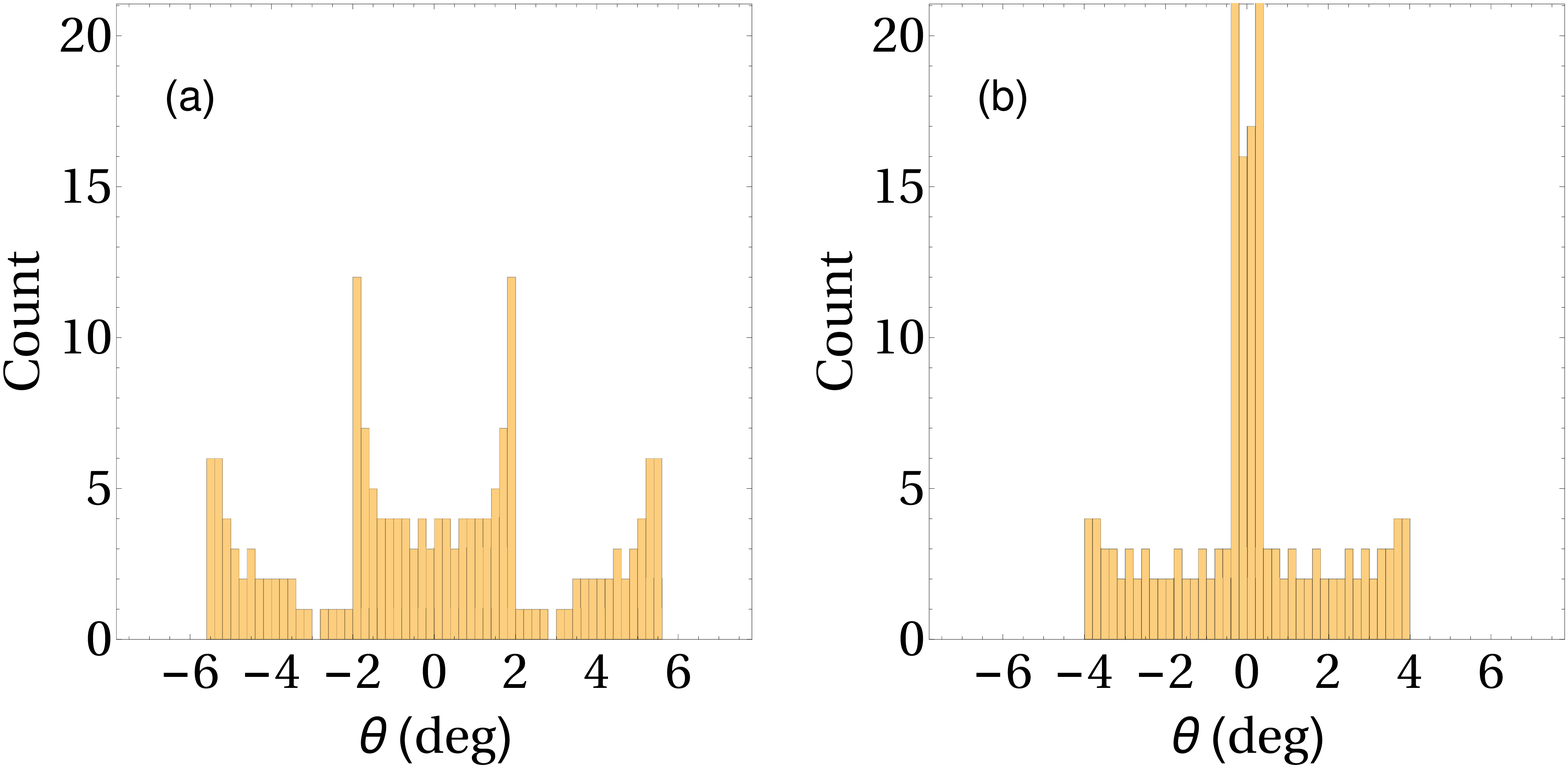}
    \end{center}
  \caption{Same as Fig.~\ref{sslit_a2_t0.09} with $\tau_v=0.0225$~s and $k_F a=2\pi$, i.e. $a=\lambda_F$.}
  \label{sslit_a1_t0.0225}
\end{figure}
It should be noticed that the range of the angular far field directions are very similar to the ones in Fig.~\ref{sslit_a2_t0.0225}. Nevertheless, there is, for higher memory, a large proportion of trajectories that keep a rectilinear motion perpendicular to the slit orientation in the far field. We finally note that all observed patterns are symmetric under $x\mapsto -x$, which is a symmetry of the whole problem.

Our model can reproduce qualitatively the diffraction patterns observed in \cite{Couder2006,Bush_PT,PhD_Harris} even though the range of the angular far field computed from our model is much smaller than in those experiments. These patterns are found to be stronger when the friction time scale $\tau_v$ is decreased. More generally, we want to stress that the diffraction pattern is expected to strongly depend also on other parameters of the experiments, such as the droplet mass. We finally note that our far-field patterns cannot be fitted by the diffraction profile derived from the optical theory of Fraunhofer. This sets a limitation to a possible analogy between a walking droplet and quantum particle.
\section{Conclusion}
\label{conclusion}

In summary we have introduced a model defined by Eqs.~(\ref{Newton_drop}) and (\ref{ansatz}) which aims to describe a walking droplet on a vibrating bath in the presence of boundaries. Our model is based on Green function approach, which is a very common tool used in the context of linear partial differential equations. It allows us to treat any geometry of the tank and any number and shape of obstacles inside it. In the case of a single slit geometry, an analytical treatment of the problem is possible, and was carried out in this paper. The resulting analytical expression for the Green function was used to numerically propagate an ensemble of trajectories across the slit. While the resulting diffraction pattern is more narrow than the ones in the experiments \cite{Couder2006,Bush_PT,PhD_Harris}, a high \old{selection}\new{selectivity} of the far-field directions of the droplet's trajectory is encountered, which is qualitatively similar to the experimental outcome. We furthermore noted a significant dependence of the diffraction profile with respect to variations of the involved parameters. 

Our model can be generalised to treat any geometry without any additional parameter. We note that our formalism applies not only to open but also to closed geometries. In the latter case it is worth stressing that for the previously considered (square, circular, annular) geometries of the tank an exact and explicit expression for the Green function can be derived. In more general cases or with other (Robin) boundary conditions the Green function will have to be computed numerically, e.g. via the boundary element method, which has been proved to be very efficient in quantum scattering problems.


\new{A particularly intriguing perspective of our work is to use more intrinsic arguments from fluid dynamics in
order to give a more fundamental justification of our model. It is especially needed for obtaining a more realistic
description of the boundary conditions at the obstacle. Furthermore a more accurate treatment of an obstacle
would require to account for capillary waves emitted at each impact. Besides, the spatial damping of the Faraday
waves might also lead to some effects which are not accounted for within our model. Finally in this study it was
also always assumed that the vertical bouncing of the droplet was synchronous with the surface profile
oscillation. It should be noted that a previous study [10] already mentioned the possibility of aperiodic vertical
motion of the droplet and this was also observed in [16].We want to stress that this is also possible within our
approach. Indeed, the bouncing and walking dynamics of a droplet can be numerically propagated in the threedimensional
space without reducing it to the effective two-dimensional Newtonian equation Eq.~(\ref{Newton_drop}), such that the
position and instant of a bounce have to be determined from the intersection of the droplet’s free-fall parabola
with the space- and time-dependent surface wave profile. As subsequent bounces are therefore not necessarily
synchronised with multiples of the shaking period, equation (7) describing the surface wave profile has to be
generalised accordingly, which in particular amounts to accounting for the periodic temporal oscillations of
each partial wave amplitude that emanates from a previous bounce. Apart from these complications, no further
technical modifications concerning the determination of the surface profile are needed in order to describe
aperiodic walkers.}

Last, we want to suggest that our approach can be straightforwardly generalised in order to simulate the dynamics of several interacting droplets. While this is a much more challenging problem, our Green function approach looks like a promising candidate in order to understand the highly complex dynamics of several walking droplets in presence of obstacles.

\ack
R.D. acknowledges fruitful discussions with J.-B. Shim and W. Struyve. M.H. acknowledges fruitful discussions with M. Labousse.
This work was financially supported by the 'Actions de Recherches Concert\'ees (ARC)' of the Belgium Wallonia Brussels Federation under contract No.~12-17/02.
Computational resources have been provided by the Consortium des \'Equipements de Calcul Intensif (C\'ECI), funded by the Fonds de la Recherche Scientifique de Belgique (F.R.S.-FNRS) under Grant No. 2.5020.11.

\appendix
\section{Derivation and evaluation of the Green function {for} the single\new{-}slit problem}
\label{Green}

We choose to look for the Green function {for} the single\new{-}slit scattering problem in a form of a series expansion. This form is more suitable for numerical implementation and is more accurate for small or moderate wavelength. As explained in the main text, it is especially useful to use elliptic coordinates in order to account for a single\new{-}slit obstacle. Indeed each arm of the slit has a very simple expression in these coordinates.
Besides, we are interested here in Neumann boundary conditions on the obstacle. This restricts the set of function, which can be used for the expansion. In the following we write an ansatz for the Green function as a series of Mathieu functions, see \ref{Mathieu} for their definition and basic properties. Then the conditions are fixed for that ansatz to actually solve Eq.~(\ref{Green_eq}) with the required boundary conditions. Our derivation follows closely the steps described in \cite{strutt}.

Start by writing an ansatz for the Green function in both half planes. Without loss of generality the source is supposed to be in the lower half plane, which means that $v_0<0$. Then the Green function can be written as:
\begin{equation}
  \label{expandG2_up}
  G({\bf r},{\bf r}_0)=
\displaystyle\sum_{n\ge 0} \alpha_n^{(+)} Me^{(1)}_n(q,u)ce_n(q,v)
\end{equation}
in the upper half plane $0<v< \pi$, and 
\begin{equation}
  \label{expandG2_down}
  G({\bf r},{\bf r}_0)=
\dfrac{2}{\pi}\displaystyle\sum_{n\ge 0} \dfrac{Me^{(1)}_n(q,u_>)Ce_n(q,u_<)ce_n(q,v_0) ce_n(q,v)}{Me^{(1)\;\prime}_n(q,0) ce_n(q,0) } +
\displaystyle\sum_{n\ge 0} \alpha_n^{(-)} Me^{(1)}_n(q,u)ce_n(q,v)\ .
\end{equation}
in the lower half plane $-\pi < v <0$.

These expansions obey Neumann boundary conditions along the slit's arms and outgoing boundary conditions when $|{\bf r}|\to\infty$. The first sum in Eq.~(\ref{expandG2_down}) has been chosen to fulfil the matching conditions at ${\bf r}={\bf r}_0$. Indeed we used the following decomposition \cite{sips}:
\begin{equation}
  \label{expandH0}
  \frac{H_0^{(1)}(k |{\bf r}-{\bf r}_0|)}{4\ic} + \frac{H_0^{(1)}(k |{\bf r}-{\bf r}'_0|)}{4\ic}=\frac{2}{\pi}
\sum_{n\ge 0} \dfrac{Me^{(1)}_n(q,u_>)Ce_n(q,u_<)ce_n(q,v_0) ce_n(q,v)}{Me^{(1)\;\prime}_n(q,0) ce_n(q,0) }\ .
\end{equation}
where ${\bf r}'_0$ stand for the image of ${\bf r}_0$ under the \old{symmetry}\new{transformation} $y\mapsto -y$.

The next step is to determine the remaining unknown coefficients $\alpha_n^{(+)}$ in Eq.~(\ref{expandG2_up}) and $\alpha_n^{(-)}$ in Eq.~(\ref{expandG2_down}).
It is achieved by requiring the continuity of both the function and its normal derivative across the slit. Recall first that the slit is described in elliptic coordinates by $u=0$ and $-\pi<v<\pi$. More precisely, the slit is seen in these coordinates as an ellipse with a unit eccentricity.
The continuity condition for $G({\bf r},{\bf r}_0)$ at the slit  reads:
\begin{equation}
  \sum_{n\ge 0} \alpha_n^{(+)} Me^{(1)}_n(q,0)ce_n(q,v)=\dfrac{2}{\pi}\displaystyle\sum_{n\ge 0} \dfrac{Me^{(1)}_n(q,u_0)}{Me^{(1)\;\prime}_n(q,0)}ce_n(q,v_0) ce_n(q,-v)+\sum_{n\ge 0} \alpha_n^{(-)} Me^{(1)}_n(q,0)ce_n(q,-v)\ . \label{cond1}
\end{equation}
The condition for the continuity of the normal derivative across the slit is:
\begin{equation}
  \sum_{n\ge 0} \alpha_n^{(+)} Me^{(1)\;\prime}_n(q,0)ce_n(q,v)=-\sum_{n\ge 0} \alpha_n^{(-)} Me^{(1)\;\prime}_n(q,0)ce_n(q,-v)\ . \label{cond2}
\end{equation}
We used that $Ce_n(q,u)=ce_n(q,\ic u)$ so $Ce_n(q,0)=ce_n(q,0)$.
It is crucial to notice that both Eqs.(\ref{cond1}) and (\ref{cond2}) are written for $0< v < \pi$. The orthogonality of the angular Mathieu functions on this restricted range
\begin{eqnarray}
  \label{ortho_ce1}  \displaystyle\int_0^\pi ce_n(q,v)ce_p(q,v)\ud v = \dfrac{\pi}{2}\delta_{n,p}\ .
\end{eqnarray}
 is used to obtain a linear system for the unknown coefficients:
\begin{eqnarray}
  \label{cond11}
  \alpha_p^{(+)} Me^{(1)}_p(q,0)&-\alpha_p^{(-)} Me^{(1)}_p(q,0)=& \dfrac{2}{\pi}\dfrac{Me^{(1)}_p(q,u_0)ce_p(q,v_0)}{Me^{(1)\;\prime}_p(q,0)} \ ,\\
\label{cond22}
\alpha_p^{(+)} Me^{(1)\;\prime}_p(q,0)&+\alpha_p^{(-)} Me^{(1)\;\prime}_p(q,0)=&0\ .
\end{eqnarray}
The determinant of this linear system is $2Me^{(1)}_p(q,0)Me^{(1)\;\prime}_p(q,0)$, hence is finite for $q>0$. The coefficients are then uniquely determined:
\begin{eqnarray}
  \label{aplus}
  \alpha_p^{(+)}&=&\dfrac{1}{\pi}\dfrac{Me^{(1)}_p(q,u_0)ce_p(q,v_0)}{Me^{(1)}_p(q,0)Me^{(1)\;\prime}_p(q,0)}\ ,\\
  \label{aminus}
  \alpha_p^{(-)}&=&-\dfrac{1}{\pi}\dfrac{Me^{(1)}_p(q,u_0)ce_p(q,v_0)}{Me^{(1)}_p(q,0)Me^{(1)\;\prime}_p(q,0)}\ .
\end{eqnarray}
Putting the expression (\ref{aplus}) back into Eq.~(\ref{expandG2_up}) on the one hand and (\ref{aminus}) into Eq.~(\ref{expandG2_down}) on the other hand gives 
Eq.~(\ref{Gsingleslit_up}) and Eq.~(\ref{Gsingleslit_down}) respectively.

The numerical evaluation of the Green function requires the computation of the Mathieu function of both first and third kinds for a large range of orders. An efficient way to evaluate these functions was to store with very high accuracy the Fourier components $A^{(n)}_{p}(q)$ defined in Eq.~(\ref{Fourier_ce_even}) and Eq.~(\ref{Fourier_ce_odd}) for $q=\pi^2$, cf. Eq.~(\ref{defq}). These coefficients were then used to evaluate $ce_n(q,v)$, $Ce_n(q,u)$ $Me^{(1)}_n(q,u)$. The radial Mathieu functions have been expanded into a series of products of Bessel functions, see e.g. \cite{McLachlan}. It is worth stressing that this common way to evaluate the radial Mathieu functions becomes rapidly inaccurate for large orders and small arguments. We then relied on a WKB$-$like approach to keep a sufficient accuracy. Technical details referring to the numerical evaluation will be provided in a forthcoming publication.

\section{Brief reminder about Mathieu functions}
\label{Mathieu}

The Mathieu functions are defined \cite{bateman3} as the solutions of the Mathieu equation:
\begin{equation}
  \label{mathieu_eq}
   y''(x)+\left[h-2q \cos(2x)\right]y(x)=0\ ,
\end{equation}
where the prime denotes differentiation with respect to $x$.
From Floquet theory Eq.~(\ref{mathieu_eq}) admits periodic solutions for a \textit{discrete} set of values of $h(q)$, called the characteristic value. For a fixed $q$ and $h=h(q)$ the periodic solution can be made real and it is usually called the Mathieu function. It is standard to distinguish between two symmetry classes:
\begin{itemize}
\item if one wants $y'(0)=0$ and $y'(\pi)=0$ then $h(q)=a_n(q)$ and the solution is denoted by $ce_n(q,v)$ for $n\ge 0$, 
\item if one wants $y(0)=0$ and $y(\pi)=0$ then $h(q)=b_n(q)$ and the solution is denoted by $se_n(q,v)$ for $n>0$.
\end{itemize}
The so obtained functions are normalised so as to form an orthogonal family:
\begin{equation}
\label{orthoM}
  \int_0^{2\pi} ce_n(q,v) ce_m(q,v)\ud v =\int_0^{2\pi} se_n(q,v) se_m(q,v)\ud v =\pi \delta_{m,n}\ ,
\end{equation}
where $\delta_{m,n}$ denotes Kronecker symbol. Last, by convention one has:
\begin{equation}
  ce_n(q,0)>0,\quad \dfrac{\ud se_n}{\ud v}(q,v)\Big|_{v=0}>0\ .\label{condA2n}
\end{equation}
In the current study we are only considering Neumann boundary condition. Therefore, we will be restricted from now on to the first symmetry class.

As any periodic function Mathieu functions can be expanded as Fourier series. It is useful to distinguish whether $n$ is odd or even:
\begin{eqnarray}
  \label{Fourier_ce_even}
  ce_{2n}(q,v)&=&\displaystyle\sum_{p=0}^\infty A^{(2n)}_{2p}(q) \cos(2 p v) ,\\
  \label{Fourier_ce_odd}
  ce_{2n+1}(q,v)&=&\displaystyle\sum_{p=0}^\infty A^{(2n+1)}_{2p+1}(q) \cos\left[(2p+1)v\right]\ .
\end{eqnarray}

In the same spirit the radial (or modified or associated) Mathieu functions are defined as solution of the radial Mathieu equation:
\begin{equation}
  \label{mod_mathieu_eq}
   y''(x)-\left[h-2q\cosh(2x)\right]y(x)=0\ .
\end{equation}
When $h$ is equal to a characteristic value $a_n(q)$, it is useful to define the following solutions of Eq.~(\ref{mod_mathieu_eq}):
\begin{eqnarray}
  \label{defCeMe}
&  h=a_n(q),&\quad y(u)=Ce_n(q,u) \textrm{  or  } y(u)=Me^{(1)}_n(q,u)\ ,
\end{eqnarray}
obeying the following constraints:
\begin{itemize}
\item $Ce_n(q,u)$ is a real even smooth solution of Eq.~(\ref{mod_mathieu_eq}) for $h=a_n(q)$,
\item $Me^{(1)}_n(q,u)$ is the only solution of Eq.~(\ref{mod_mathieu_eq}) obeying Sommerfeld's radiation condition at infinity for $h=a_n(q)$ and such that $\re Me^{(1)}_n(q,u)=Ce_n(q,u)$.
\end{itemize}
Notice that one has $Ce_n(q,u)=ce_n(q,\ic u)$.
The functions $Ce_n,Me^{(1)}_n$ can be shown to be linearly independent. They can be used to expand any solution of Eq.~(\ref{mod_mathieu_eq}) when $h=a_n(q)$.

\end{document}